\documentclass{article}
\usepackage{spconf,amsmath,graphicx}
\usepackage{amssymb} 
\usepackage{booktabs}

\title{
    Learning accurate rigid registration \\
    for longitudinal brain MRI from synthetic data
}

\name{
    Jingru Fu$^{1}$,
    Adrian V.\ Dalca$^{2\text{--}5}$,
    Bruce Fischl$^{2\text{--}4}$,
    Rodrigo Moreno$^{1}$,
    Malte Hoffmann$^{2\text{--}4}$
}

\address{
    $^1$Division of Biomedical Imaging, KTH Royal Institute of Technology, Huddinge, Sweden \\
    $^2$Athinoula A.\ Martinos Center for Biomedical Imaging, Charlestown, USA \\
    $^3$Department of Radiology, Massachusetts General Hospital, Boston, USA \\
    $^4$Department of Radiology, Harvard Medical School, Boston, USA \\
    $^5$Computer Science \& Artificial Intelligence Laboratory, MIT, Cambridge, USA \\
}

\begin{document}

\maketitle

\begin{abstract}
Rigid registration aims to determine the translations and rotations necessary to align features in a pair of images. While recent machine learning methods have become state-of-the-art for linear and deformable registration across subjects, they have demonstrated limitations when applied to longitudinal (within-subject) registration, where achieving precise alignment is critical. 
Building on an existing framework for anatomy-aware, acquisition-agnostic affine registration, we propose a model optimized for longitudinal, rigid brain registration. By training the model with synthetic within-subject pairs augmented with rigid and subtle nonlinear transforms, the model estimates more accurate rigid transforms than previous cross-subject networks and performs robustly on longitudinal registration pairs within and across magnetic resonance imaging (MRI) contrasts.
\end{abstract}


\section{Introduction}
\label{sec:intro}

\begin{figure}[!t]
\centering
\centerline{\includegraphics[width=8.9cm]{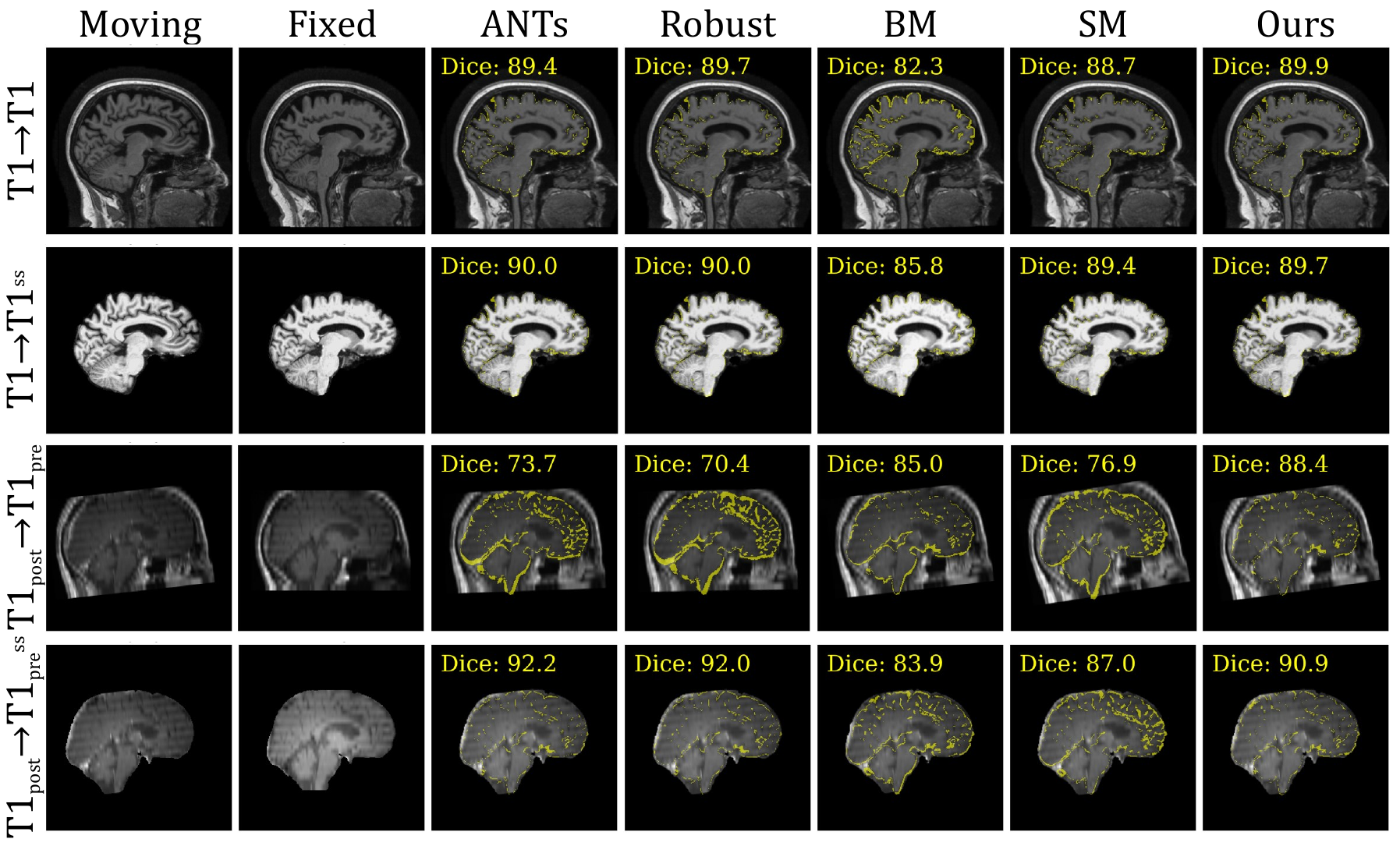}}
\caption{Representative within-subject registration pairs. We overlay the image moved by each method with the absolute difference between fixed and moved brain masks in yellow. BrainMorph (BM) and SynthMorph (SM) use deep learning. 
}
\label{fig:example}
\end{figure}

Linear image registration is a cornerstone of medical neuroimage processing and analysis that seeks to find a global spatial transformation that optimally aligns structures between two images. Common use cases include aligning brains \textit{across} different subjects to normalize anatomical variability, initializing study-specific brain template construction \cite{fischl2002whole,evans2012brain}, aligning brains \textit{within} the same subject at different time points to remove positional differences in longitudinal studies \cite{reuter2010highly,takao2012longitudinal}, and to fuse information from different modalities~\cite{chandrashekhar2021cloudreg,han2022review}. 

Linear registration is also typically a first step before deformable image registration (DIR) \cite{balakrishnan2019voxelmorph,hoffmann2021synthmorph}.
While classical algorithms can achieve high accuracy, they require time-consuming optimization for every new image pair. For certain applications that demand real-time processing, such as prospective motion correction during neuroimaging scans, sub-second registration is necessary \cite{tisdall2012volumetric}. Moreover, most classical methods require skull stripping to remove irrelevant content that moves independently of the brain and can impinge on the accuracy of brain-specific registration. 
Deep-learning (DL) based approaches are promising to tackle these two issues. 
Recent DL methods predict keypoints between images from different subjects that can be used to obtain the optimal rigid or affine transform via a differentiable closed-form expression~\cite{moyer2021equivariant,evan2022keymorph,hoffmann2023anatomy}. 
 
In this paper, we show that models designed for cross-subject registration often fail to achieve accurate within-subject registration, as these tasks have a different focus. Specifically, within-subject registration seeks precise rigid transforms to address changes in head position, scanner setup, and subtle biological variation over time. Cross-subject models, however, are trained to handle larger anatomical differences between subjects, shifting the focus from capturing subtle within-subject differences. Consequently, minor alignment errors can lead to matching voxels of different tissue types across time, potentially resulting in incorrect interpretation of tissue changes or pathology.  
We show that existing affine registration models may not capture the precise rigid transformations needed for accurate alignment, as they include scaling and shearing components that are not necessary (and indeed incorrect!) for rigid motion \cite{hoffmann2024anatomy}. 
We develop an accurate DL model for within-subject rigid registration.

\noindent\textbf{Synthesis strategy.} Obtaining sufficient longitudinal training data for learning-based methods is more challenging than for registration across subjects due to the limited availability of image pairs from the same subject across different time points or modalities. Recent advances in learning strategies allow neural networks to be trained without acquired images \cite{hoffmann2021synthmorph,billot2023synthseg,hoopes2022synthstrip, iglesias2023synthsr}, by generating a virtually endless stream of synthetic training data. SynthMorph \cite{hoffmann2023anatomy} leverages synthetic data generation for robust cross-subject affine registration, providing flexibility in forming training pairs. By focusing the loss function on targeted anatomical labels, which are independent of image contrast by design, the approach leads to models that are anatomy-aware and acquisition-agnostic.

\section{Methods}
\label{sec:methods}

\noindent\textbf{Model architecture.}
We use a fully convolutional feature detector (referred to as \textit{Detector} hereafter)~\cite{moyer2021equivariant,evan2022keymorph,hoffmann2023anatomy}, which predicts $k$ ReLU-activated feature maps for each of the input images $\{m, f\}$. From these, we compute moving and fixed barycenters $\{a_i\}$, $\{b_i\}$, and associated weights $\{p_i\}$ and $\{q_i\}$ (sum normalized over $k$ channels) for channel $i \in \{1,2,...,k\}$. We fit a rigid transform $T_{\theta}$ via a closed-form weighted least-squares expression \cite{moyer2021equivariant} solving:
\begin{equation}
    T_\theta = \arg\min_{t} \sum_{i=1}^k w_i \left\| a_i^\top - \begin{bmatrix} b_i^\top & 1 \end{bmatrix} t^\top \right\|^2,
\end{equation}
with weights $w_i =p_i q_i$.
We use a truncated U-Net architecture as the \textit{Detector} backbone as proposed in \cite{ulyanov2016instance, wang2024brainmorph}, consisting of ten 3D convolutional blocks. The encoder applies $w=[32, 64, 128, 256, 512, 1024]$ and the decoder $w=[512, 256, 128, 64]$ filters. An eleventh convolutional block outputs $k=256$ feature maps. We downsample the network and loss inputs by a factor of 2. By using a deeper network than affine SynthMorph \cite{hoffmann2023anatomy}, we increase the model capacity to account for subtle misalignment within subjects.

\noindent\textbf{Synthetic training data.}
Instead of training with real images, we synthesize variable data at each training iteration from brain label maps with an image generator \cite{hoffmann2021synthmorph,hoffmann2023anatomy,hoffmann2024anatomy}. This strategy enables multi-contrast registration: 
\textit{(i)} We randomly select a pair of moving and fixed brain label maps $\{s_m, s_f\}$; \textit{(ii)} we spatially augment $\{s_m, s_f\}$ by applying the composition of a random affine and nonlinear transform; \textit{(iii)} we sample a mean intensity and assign it to all voxels associated with each label $j \in K$ in $s_f$, and separately for $s_m$, where $K$ is the set of all labels, yielding a gray-scale image pair; \textit{(iv)} finally, we corrupt the synthesized images as in prior work \cite{hoffmann2023anatomy} (see reference for examples).

Our synthetic generator has three key differences compared to prior work \cite{hoffmann2023anatomy}. 
\textit{(i)} We construct the moving and fixed label maps from a single input label map to build an intra-subject pair at each iteration; \textit{(ii)} we remove scaling and shear from the spatial augmentation, as these are out of distribution for rigid registration; \textit{(iii)} we augment with more subtle nonlinear transforms than those used for cross-subject registration to accommodate the lower variability in longitudinal data, by adjusting the strength and smoothness of the deformation (Section~\ref{sec:exper}). We create the nonlinear transform by integrating a randomly sampled and smoothed stationary velocity field, simulating nonlinear temporal effects such as differential gradient nonlinearities, B0 distortion and nonrigid motion of anatomy such as the eyes \cite{hoffmann2021synthmorph}.

\noindent\textbf{Loss function.}
To encourage the network to register specific anatomy while ignoring irrelevant image content, we re-label $\{s_m, s_f\}$ to only include anatomical labels $J$ consisting of larger tissue classes (training details). We optimize the parameters $h_{\theta}$ of \textit{Detector} using a mean squared error (MSE) loss $\mathcal{L}$ on one-hot encoded label maps.
Ensuring inverse consistency in within-subject registration methods is crucial to avoid introducing bias in longitudinal studies \cite{reuter2010highly,reuter2012within}. We maintain inverse consistency by resampling both labels $\{s_m, s_f\}$ into a halfway space during training. That is, instead of mapping $s_m$ onto $s_f$, we compute $s_m\circ T_{\theta}^{1/2}$ and $s_f\circ T_{\theta}^{-1/2}$ to apply the same amount of interpolation to each label map:
\begin{equation} 
    \mathcal{L} = \frac{1}{|\Omega|}\sum_{\substack{j\in J \\ x \in \Omega}}[(s_m|_j\circ T_{\theta}^{1/2})(x)-(s_f|_j\circ T_{\theta}^{-1/2})(x)]^2,
\end{equation}
where $s|_j$ represents the one-hot encoded label $j \in J$ of label map $s$ defined at the voxel locations $x\in \Omega$, and the transform $T_{\theta}$ maps the discrete spatial domain $\Omega$ of $s_f$ onto $s_m$.

\noindent\textbf{Training details.}
We train the model using the ADAM optimizer with a learning rate of $10^{-5}$ and batch size one until the loss on the validation set plateaus. For fast convergence, we optimize the overlap of $|J|=3$ merged brain-tissue classes (left, right hemisphere and cerebellum) at the beginning of training. We switch to $|J|=5$ finer classes (including left and right cortex and subcortical gran matter, and cerebellum) at later stages of the training. Training with a higher number of classes $|J|$ does not improve performance.

\begin{figure*}[t!]
    \centering
    \begin{minipage}[b]{.49\linewidth}
        \centering
        \includegraphics[width=\linewidth]{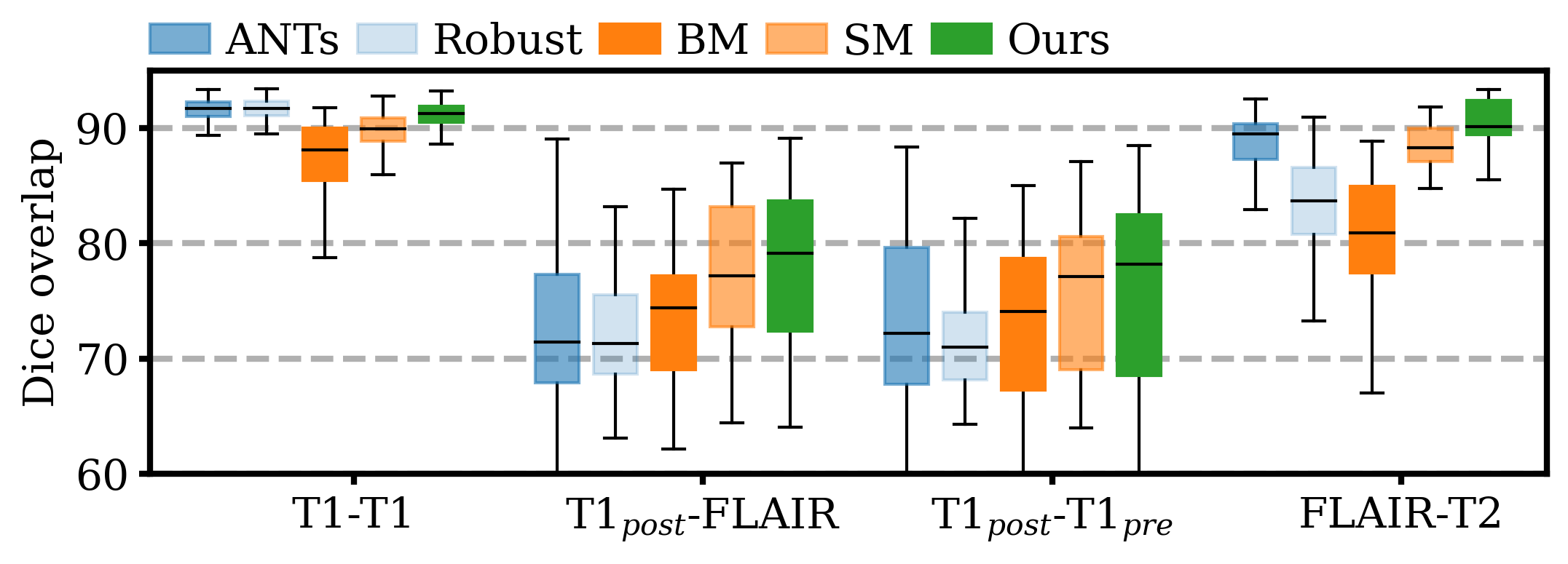}
        \centerline{Whole head}\medskip
    \end{minipage}
    \hfill
    \begin{minipage}[b]{.49\linewidth}
        \centering
        \includegraphics[width=0.94\linewidth]{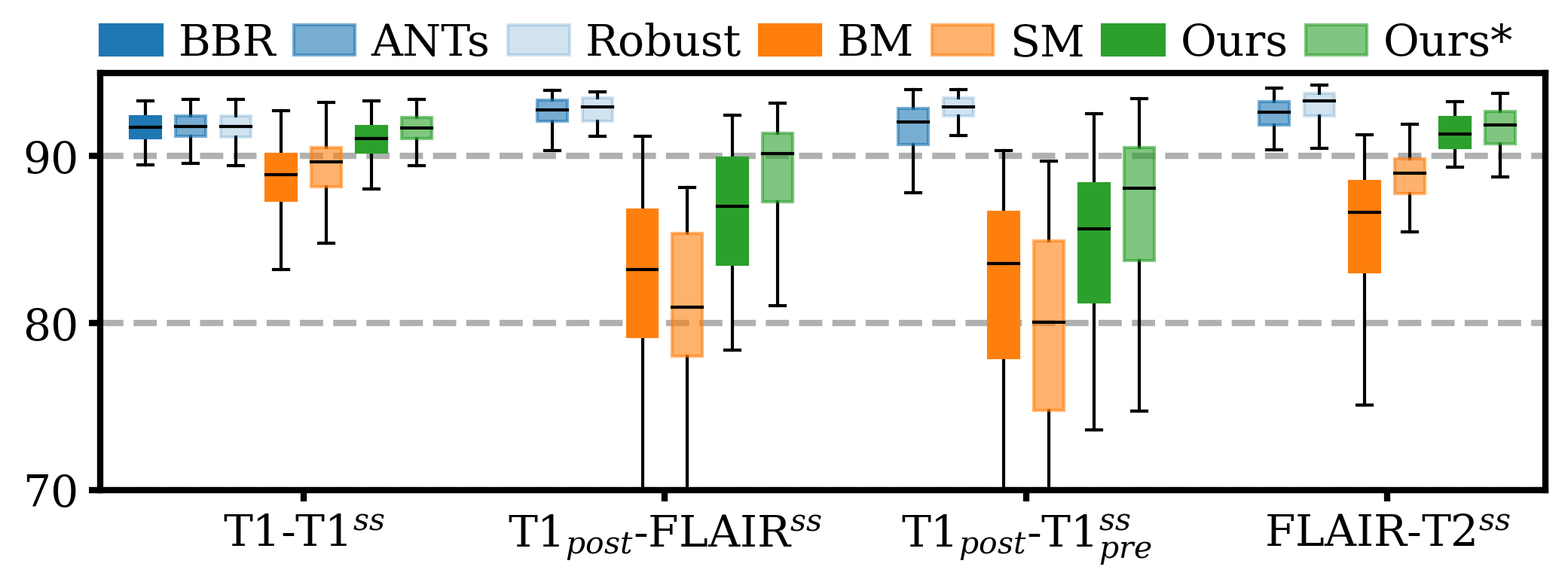}
        \centerline{Skull-stripped}\medskip
    \end{minipage}
    \caption{Rigid 3D registration accuracy on the test set, as mean Dice scores over left and right cortex, subcortex, and cerebellum. bbregister (BBR) uses FreeSurfer reconstructions, which include brain masks, and is thus classified as skull-stripped. The asterisk denotes instance-specific optimization after inference.}
    \label{fig:res}
\end{figure*}

\noindent\textbf{Instance-specific optimization.} Instance-specific optimization can refine the initial deformation field generated by a deep learning model \cite{balakrishnan2019voxelmorph}. We adapt this approach for rigid registration by fine-tuning the transform predicted by our model using gradient descent over 200 iterations for each test pair. We optimize MSE within and mutual information (MI) across contrasts. This optimization is most effective for skull-stripped images, as non-brain regions in whole-head images can lead to misalignment of the anatomy of interest due to irrelevant tissue or background differences.

\section{Experiments and results}
\label{sec:exper}

\textbf{Datasets.} For synthesizing training data, we use the same 100 whole-head and skull-stripped tissue segmentations as SynthMorph \cite{hoffmann2024anatomy}. Our evaluation set includes T1w brain scans from the longitudinal cohorts ADNI \cite{jack2008alzheimer} (at 1.5T and 3T) and MIRIAD \cite{malone2013miriad}. The evaluation set also includes QIN \cite{Mamonov2016} scans with T2w, pre and post-gadolinium T1w (T1w$_{pre/post}$), and FLAIR stacks of axial 5-mm slices. QIN is a clinical dataset of subjects with newly diagnosed glioblastoma. We compare FreeSurfer-processed whole-head and skull-stripped images for ADNI and MIRIAD. For the QIN dataset, we use brain labels obtained from isotropic T2-SPACE scans to mask out non-brain regions in other sequences from the same acquisition session. We conform all images to the same isotropic $256\times256\times256$ 1-mm voxel resolution with left-inferior-anterior (LIA) orientation. We derive brain labels for training and evaluation using SynthSeg \cite{billot2023synthseg}. 

\noindent\textbf{Competing methods.} We compare the model with the following methods. \textit{(i)} Robust \cite{reuter2010highly}. For within-contrast registration, we use the robust cost function, which can down-weight the contribution of local change in the image. 
For cross-contrast registration, we choose the experimental robust-entropy metric. \textit{(ii)} ANTs \cite{avants2011reproducible}, using the recommended parameters; \textit{(iii)} bbregister (BBR) \cite{greve2009accurate}, which requires one anatomical scan processed with FreeSurfer, limiting its applicability to ADNI and MIRIAD; \textit{(iv)} Brainmorph (BM) \cite{wang2024brainmorph}. 
We choose the best-performing, large BM network with 512 feature maps. \textit{(v)} affine SynthMorph (SM) \cite{hoffmann2023anatomy}, which focuses on cross-subject registration. All baselines other than SM directly estimate a rigid transform.

\begin{table}[t]
\caption{Acquired test (Te.) and validation (Va.) data for within-subject registration. ADNI has 4 scans per subject spanning 2 years at 0.5--1 year intervals, while MIRIAD has 7 scans per subject with intervals of 2 weeks to 2 years. QIN has 2 scans per subject per contrast acquired before and after surgery, 2--5 days apart. 
}
\begin{tabular}{llccc}
\toprule
Dataset & Contrast & Subjects & Va.\ Pairs & Te.\ Pairs \\ \midrule
ADNI-1.5T & T1w & 100 & 30 & 190 \\
ADNI-3T   & T1w & 50  & 30 & 90  \\
MIRIAD    & T1w & 40  & 60 & 70  \\
QIN       & \textit{Various} & 50 & --  & 356 \\ \bottomrule
\end{tabular}
\label{tab:dataset}
\end{table}

\noindent\textbf{Experimental setup.}
We randomly draw 5 subjects from ADNI and MIRIAD, respectively, to construct validation pairs. We use the remaining subjects along with the QIN dataset for testing (Table~\ref{tab:dataset}). To construct same-contrast test pairs, we randomly select one pair of scans from different time points for each ADNI and MIRIAD subject, either with or without skull stripping. For cross-contrast registration, we select pairs between the two available QIN sessions: T1w$_{pre}$ and FLAIR, T1w$_{pre}$ and T1w$_{post}$, and FLAIR and T2w. We evaluate registration accuracy by computing the mean Dice overlap of the moved and the fixed label maps across the finer $|J|=5$ tissue classes. 

We analyze the performance of our model in two experiments. First, we assess the accuracy of each tool on skull-stripped and whole-head image pairs, respectively. For skull-stripped images, we also report enhanced results achieved with instance-specific optimization.

Second, we analyze performance across a range of deformation strengths and levels of smoothness for the nonlinear deformation synthesis during training. In this experiment, we initialize encoder networks using cross-subject SM weights and train each parameter setting for 1M batches. We report results for the validation set.

\noindent\textbf{Results.} Fig.~\ref{fig:res} shows that the proposed model, trained with synthetic data only, consistently outperforms BM. In contrast, BM trained with over 100k real images, whereas our model trains using only 100 label maps. For whole-head images, our model achieves higher median Dice scores across all contrast pairings, except for within-contrast T1w-T1w pairs where classical optimization-based methods (shown in blue) perform slightly better ($\sim$0.4 Dice points). However, with instance-specific optimization, our model reaches the performance level of the classical methods for skull-stripped T1w-T1w pairs. As expected, for skull-stripped within-subject registration, the classical baselines BBR, ANTs and Robust consistently outperform learning-based methods (shown in orange and green) across contrasts. However, the classical methods struggle without skull stripping at cross-contrast registration; for example, ANTs' performance drops by a median of 21.4 Dice points for T1$_{post}$-FLAIR, as the algorithm cannot distinguish between brain and non-brain tissue across different contrasts. For this pairing, the performance of our model only drops by 7.9 points, likely because \textit{Detector} \textit{learns} to focus on brain structures to estimate the rigid transform. Across all experiments, our model shows superior performance and robustness relative to the tested DL models. 

\begin{figure}
\begin{minipage}[b]{\linewidth}
  \centering
  \centerline{\includegraphics[width=9.0cm]{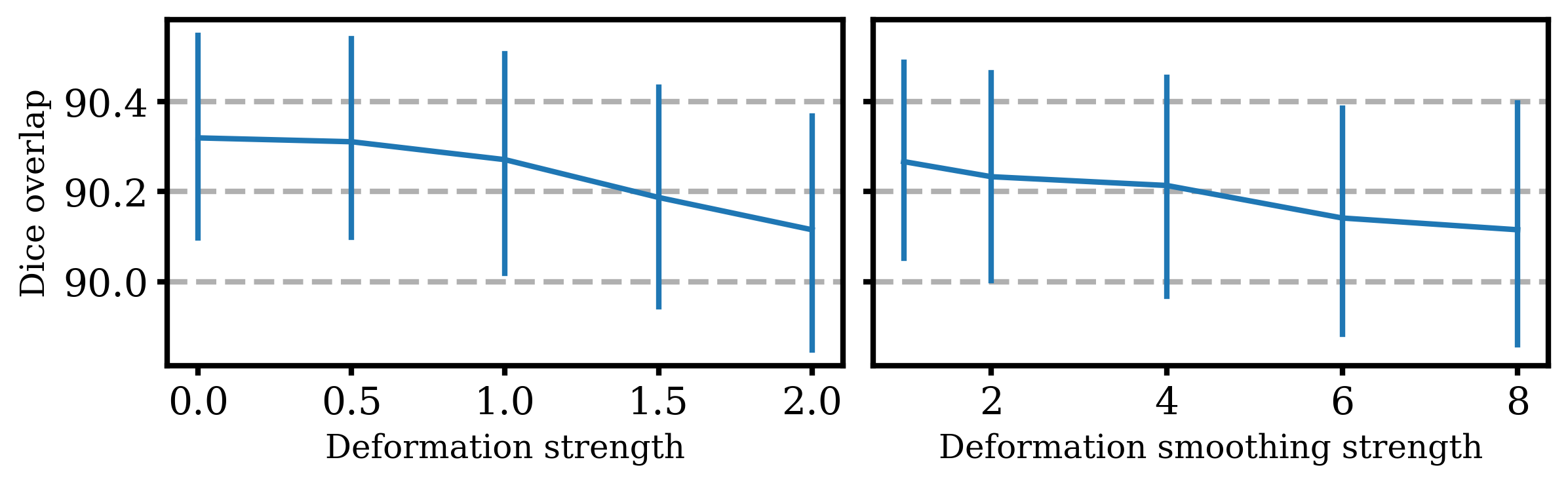}}
\end{minipage}
\caption{Effect of deformation and smoothing strengths of the non-linear deformation augmentation on registration accuracy in the validation set. Error bars indicate 95$\%$ confidence intervals (CI).}
\label{fig:abl}
\end{figure}

Fig.~\ref{fig:abl} shows the parameter sweep for deformation and smoothing strengths used for generating nonlinear deformations at training. Our model performs best for low deformation strengths between 0 and 0.5. This result is likely due to the small within-subject relative to inter-subject variation. A similar pattern emerges for the level of smoothness of the nonlinear deformation, with smaller values yielding better performance. To accommodate differential MR distortions in clinical data that, in contrast to the ADNI and MIRIAD validation set, are generally not being acquired with harmonized protocols and hardware, we chose a deformation strength of 0.5 and smoothness level of 1 to generate training data for the experiments of Fig. \ref{fig:res}.

Fig.~\ref{fig:example} presents a qualitative comparison of ADNI-3T and QIN registration pairs with and without skull-stripping.
For T1w-T1w pairs, all methods achieve similar Dice scores, likely due to the clear anatomical boundaries and high resolution that facilitate alignment within the same imaging contrast. The observed differences primarily occur at the edges of the gray matter, likely caused by minor segmentation errors. For the more challenging cross-contrast pairs, most competing methods struggle without skull-stripping. SM appears to introduce scaling and shear for whole-head pairs---it has been optimized for affine registration.

\section{Discussion}
\label{sec:discu}

We introduce a rigid registration tool for within-subject brain scans without the need for skull stripping, capable of handling various resolutions and contrasts. This tool demonstrates robust performance compared to specialized baseline methods on both high-resolution T1w longitudinal datasets and clinical datasets with varied slice thickness. 

We adapt the synthesis strategy by removing out-of-distribution spatial augmentation to enhance performance for within-subject rigid registration. While the improved model only surpasses affine SynthMorph by about 1.8 Dice points for within-contrast whole-head pairs, it achieves a 3--8 improvement in Dice for cross-contrast whole-head pairs. This disparity also highlights the need for learning-based methods developed and tested for cross-subject registration, to be rethought for within-subject applications. In such cases, where close-to-perfect alignment is often possible and spatial differences are typically small, existing learning-based methods underperform without modification, potentially reducing our ability to detect the subtle atrophy that occurs in the early stages of various neurological disease processes.

In addition to refining the synthesis strategy, we optimize the \textit{Detector} architecture to better account for subtle misalignment in within-subject pairs. Future work will explore the effect of varying network capacity on within-subject performance. Additionally, we aim to test models that take both moving and fixed images as inputs, which may outperform the current architecture that processes only one image at a time and relies on detecting robust features for the final estimation of the transformation. As robustness is an important criterion for rigid registration, we also plan to quantify successful vs failed registrations as in prior work \cite{hachicha2023robust}.

\section{Compliance with ethical standards}
\label{sec:ethics}

We signed data use agreements for access to the publicly available ADNI and MIRIAD datasets. The retrospective analysis of QIN data required no ethical approval.

\section{Acknowledgments}
\label{sec:acknowledgments}

The project was supported in part by the Swedish Childhood Cancer Foundation MT2019-0019, MT2022-0008, China Scholarship Council, Digital Futures, project dBrain, Swedish Research Council 2022-03389, MedTechLabs, mobility grant from WASP, 
NIH grants NIBIB P41 EB015896, U01 NS132181, UM1 NS132358, R01 EB023281, R01 EB033773, R21 EB018907, R01 EB019956, P41 EB030006, NICHD R00 HD101553, R01 HD109436, R21 HD106038, R01 HD102616, R01 HD085813, and R01 HD093578, NIA R56 AG064027, R21 AG082082, R01 AG016495, R01 AG070988, NIMH RF1 MH121885, RF1 MH123195, UM1 MH130981, NINDS R01 NS070963, R01 NS083534, R01 NS105820, U24 NS135561, SIG S10 RR023401, S10 RR019307, S10 RR023043, BICCN U01 MH117023, and Blueprint for Neuroscience Research U01 MH093765. The project benefited from computational hardware provided by the Massachusetts Life Sciences Center. MH maintains a consulting relationship with Neuro42, Inc. BF and AVD are advisors to DeepHealth. Their interests are reviewed and managed by Massachusetts General Hospital and Mass General Brigham in accordance with their conflict-of-interest policies.

\bibliographystyle{IEEEbib}
\bibliography{main}

\end{document}